# Weyl Phonons: The connection of topology and chirality

Tiantian Zhang[1,*], Shuichi Murakami[2,*], Hu Miao[3,*]

[1]Institute of Theoretical Physics, Chinese Academy of Sciences, Beijing 100190, China
[2]Department of Physics, Institute of Science Tokyo, Okayama, Meguro-ku, Tokyo, Japan
[3]Materials Science and Technology Division, Oak Ridge National Laboratory, Oak Ridge, TN, USA.
Correspondence should be addressed to:
ttzhang@itp.ac.cn; murakami@stat.phys.titech.ac.jp; miaoh@ornl.gov

**Topology and chirality of fermionic quasiparticles have enabled exciting discoveries, including quantum anomalous Hall liquids and topological superconductivity. Recently, topological and chiral phonons emerge as new and fast-evolving research directions. While these concepts are separately developed, they are intimately connected in the context of Weyl phonons. The couplings between chiral and topological phonons with various electronic and magnetic quasiparticles are predicted to give rise to new quantum states and giant magnetism with fundamental and applicational interests, ranging from quantum information science to dark matter detectors.**

Unlike electrons, phonons are charge neutral, spin zero and orbital free, which makes the means of modulating phonons very limited. Thus, introducing both topological and chiral degrees of freedom is helpful and vital to understand phonon-involved physical processes and applications. Topological phonons are novel collective lattice excitations that carry non-trivial topological invariants and pseudo-spin tectures[1-3], as shown in Fig. 1. Chiral phonons refer to phonon modes that have finite angular momentum (AM) and widely present in non-centrosymmetric materials[4-8]. In this comment, we overview theoretical understandings of topological and chiral phonons and elucidate the fundamental connection between topology and chirality in the context of Weyl phonons. We discuss recent experimental progresses related to topological and chiral phonons and open questions and future research directions.

## An overview of topological and chiral phonons

**Topological phonons**

Like topological electronic quasiparticles, topological phonons are characterized by topological invariants that are intimately related to crystalline symmetries. For instance, topological nodal-line phonons, which are characterized by the geometrical Berry phase, can emerge when crystals preserve mirror ($\mathcal{R}$)/inversion ($\mathcal{P}$) symmetry[1]. Weyl phonons, on the other hand, emerges in non-centrosymmetric structures and are described by the Chern number ($C$)[2,3]. As an example, we considering the following effective two-orbital Hamiltonian:

$$H(\boldsymbol{q}) = \begin{pmatrix} Aq_xq_yq_z & B^*(q_x^2+\omega^2 q_y^2 + \omega q_z^2) \\ B(q_x^2+\omega^2 q_z^2 + \omega q_y^2) & -Aq_xq_yq_z \end{pmatrix}, \quad (1)$$

where $\omega = e^{\frac{2\pi i}{3}}$, $A$ is a real constant and $B$ is a complex constant. The $q_{x/y/z}$ are pseudo-momentum of phonons. This Hamiltonian is defined as a square root of the dynamical matrix, with its eigenvalues $\psi_\pm$ being the phonon polarization vector, $\boldsymbol{\varepsilon}_{\boldsymbol{q},\pm}$, and its eigenvalues being the frequencies of the phonons. Equation (1) describes the lattice dynamics in the vicinity of a twofold degenerate high symmetry point that respect chiral cubic symmetries and time-reversal symmetry (see Fig. 1). The topological number $C = \pm 4$ can then be derived from the eigen vectors, $\psi_\pm$, of Eq. (1) and represent twofold quadruple Weyl phonons[3,4].

**Chiral phonons**

Chiral phonons, also referred to as circularly polarized phonons, initially mean phonons with nonzero AM[5-6], $\boldsymbol{l}_{\boldsymbol{q},\sigma}$:

$$\boldsymbol{l}_{\boldsymbol{q},\sigma} = (\boldsymbol{\varepsilon}_{\boldsymbol{q},\sigma}^\dagger M \boldsymbol{\varepsilon}_{\boldsymbol{q},\sigma})\hbar, \quad (2)$$

where $\boldsymbol{\varepsilon}_{\boldsymbol{q},\sigma}$ is the polarization vector for phonon mode $\sigma$ at momentum $\boldsymbol{q}$, $(M_i)_{jk} = (-i)\varepsilon_{ijk} \otimes I_{n\times n}$ ($i,j,k = x,y,z$) is a product of the generator of SO(3) rotation and the unit matrix for a unit cell with $n$ atoms, and $\varepsilon_{ijk}$ is the Levi-Civita symbol. From a symmetry point of view, chiral phonons with finite AM are widely present in crystals, unless certain symmetries restrict the AM to be zero, such as mirror, spatial inversion, or time-reversal symmetry.

**Phonon modes with $\hat{C}_n$-symmetry**

Although AM is a fundamentally important concept with significant macroscopic consequences, it is usually not conserved in microscopic quasiparticle scattering processes. Instead, the pseudo-angular momentum (PAM), $l_q^{ph}$, has been introduced in systems with rotational or screw rotational symmetries[7-11]:

$$\hat{C}_n \boldsymbol{u}_q = e^{-\frac{2\pi i}{n} \cdot l_q^{ph}} \boldsymbol{u}_q, \quad (3)$$

$$\boldsymbol{u}_q = \boldsymbol{\varepsilon}_q e^{i(\boldsymbol{R}_l \cdot \boldsymbol{q} - \omega t)}, \quad (4)$$

where $\hat{C}_n$ is the rotation or screw rotation operator and $\boldsymbol{u}_q$ is the phonon Bloch wave function at $\boldsymbol{q}$. Since $l_q^{ph}$ originates from phase factors acquired by discrete $\hat{C}_n$ symmetry, the phonon PAM can only be defined at rotation-invariant momenta and it is conserved modulo $n$.

**Connections between topological and chiral phonons**

Topological and Weyl phonons are prevalent in materials, particularly in chiral crystals. Symmetries of the system constrain phonon modes, thereby determining topological and chiral properties. Here we use Weyl phonons to draw the connection between these independently developed fields. Consider Eq. (1) as an example, the eigen vector $\psi_\pm$ can be analytically derived in the twofold quadruple Weyl system BaPtGe that features chiral lattice motions[3-4], as shown in Fig. 1. Indeed, this connection is general for non-centrosymmetric materials, where topological phonons are Weyl phonons with nonzero AM. Namely, Weyl phonons are characterized by the Chern number, which is an integral of the Berry curvature. Since the Berry curvature has the same symmetry with the phonon AM, nonzero Chern number implies that the phonon AM should be nonzero around the Weyl point. Therefore, Weyl phonons are a special type of chiral phonons, as highlighted in Fig. 1. Since phonons can be chiral even away from point or line degeneracies, chiral phonons are not necessarily topological.

Furthermore, it has been shown that[3] Weyl phonons can be classified based on the (screw) rotational symmetries $\hat{C}_{n=3,4,6}$, which also determines the phonon PAM (see Eq. (3)). Therefore, the Chern numbers $C$ of Weyl phonons can be derived from the PAM of the degenerated phonon modes[12-13]. This mapping is, however, symmetry dependent.

**Experimental evidence of topological phonons and chiral phonons**

**Topological phonons were first revealed using** inelastic x-ray scattering (IXS)[14]. By quantitatively comparing the IXS determined phonon dynamical structure factor, $S(\boldsymbol{Q}, \omega)$, and density functional theory calculated $S(\boldsymbol{Q}, \omega)$, the double Weyl phonons, hence chiral phonons as described in previous section, are established in the $\mathcal{P}$-breaking crystal FeSi[14]. Observations of truly chiral phonons were also reported in Raman and resonant inelastic x-ray scatterings[7-8, 14-15]. The topologically trivial chiral phonons in 2D transition metal dichalcogenides $WSe_2$ were first reported by using the circular dichroism (CD) in the transient infrared spectroscopy[16].

**Outlook**

**1. Topological edge modes**

Topological phonons can give rise to edge modes, such as the predicted double helicoidal surface phonon modes in transition metal monosilicides, and the flat surface phonon modes in $MoB_2$ and high-$T$c conventional superconductor $MgB_2$. Experimental observation of these novel topological edge modes will be important for the understanding of topological bosonic excitations and related physical consequences. When the time-reversal symmetry is broken, quantum Hall analogous edge modes can be obtained[17]. These topologically protected edge modes can be potentially applied to thermal diodes, thermal transistors, and other thermal devices.

**2. Phonon angular momentum**

The AM of chiral phonons has been proposed for potential applications in quantum information science and microelectronics, such as phonon Einstein-de Haas effect[18], which can transduce thermal energy to rigid-body kinetic energy and serve as thermo-motors and thermo-switch, and also new approach for high-precision dark matter detectors via interacting with dark matters[19]. Chiral phonon also drives novel quantum states and phenomena, such as transverse Peierls transition[20], axionic charge density waves[21], and superconductivity[22], spin Seeback effect[23], giant phonon magnetism[24-26], etc.

**3. Phonons that are both topological and chiral**

When a phonon exhibits characteristics of being both topological and chiral, its applications also encompass those in separate fields. For example, in Weyl phonon systems, the chiral fermions can also appear, based on the electronic wavefunctions at the Fermi surface. Thus, the phonon modes can also couple with chiral fermions, such as CoSi, and give rise to non-linear Hall effect or novel spectroscopic response. Exploration on the Weyl phonon-triggered unusual quantum states will be a highly interesting research direction. Previous studies have showed the relationship between the pseudospin and the Chern number[3-4]. The understanding of the connection between spin/orbital AM texture and topology of phonons may open a new route to control and manipulate topological and chiral phonons.

**Conclusion**

Topological and chiral phonons are fast-developing research fields with intriguing potential for both fundamental research and applications. Beyond the experimental verifications, the exploration of topological and chiral phonon-driven quantum phenomena is an exciting research frontier, such as chiral charge density waves[27], ultrafast demagnetization[28], etc. While numerous novel physical properties have been reported, comprehensive understanding of novel phenomena, such as giant phonon magnetism and thermal Hall effect, remain elusive, presenting a fundamental obstacle for further applications. The integration of theory, electronic structure databases [29-31], phonon databases[32-33], and machine learning techniques may offer a promising avenue for overcoming those fundamental challenges.

**Acknowledgements**

T. Z. acknowledges the support from National Key R&D Project (Grant Nos. 2023YFA1407400 and 2024YFA1400036) and National Natural Science Foundation of China (Grant Nos. 12374165 and 12047503). S.M. is supported by JSPS KAKENHI Grant Nos. JP22H00108 and JP24H02231. H. M. is supported by the U.S. Department of Energy, Office of Science, Basic Energy Sciences, Materials Sciences and Engineering Division. Figure 1 is adapted from https://doi.org/10.1103/PhysRevB.103.184301.

**Competing interests**

The Authors declare no competing interests.

**Author Contributions**

T. Z., S. M. and H. M. devised the project idea and prepared the manuscript.

**Figures**

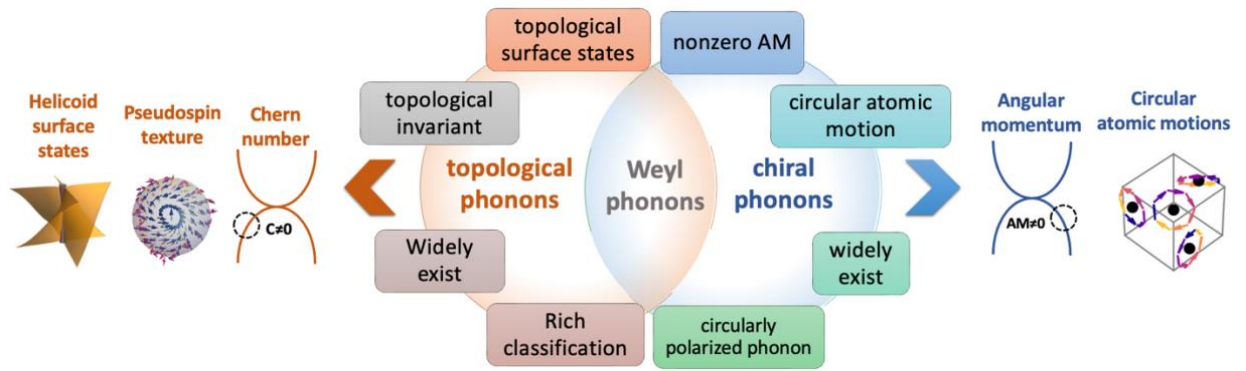

**Figure 1: Chiral and topological phonons in quantum materials.** Topological phonons are characterized by non-trivial topological invariants, associated with nontrivial pseudospin texture around the band degeneracies and topological surface states[2-3]. Chiral phonons are described by phonon modes with finite AM, associated with circular atomic motions in the real space, thus also referred to as circularly polarized phonons. These two concepts are fundamentally

connected in 3D non-centrosymmetric materials. Left panels show the spiral surface state, the pseudo-spin texture, and band dispersion of the twofold quadruple Weyl phonon with C=+4. Right panels show that in the vicinity of twofold quadruple Weyl point, the corresponding lattice motion is chiral with finite AM[3-4]. The general relation between Weyl and chiral phonons are highlighted in the middle panel. In non-centrosymmetric materials, Weyl phonons are special type of chiral phonons. Chiral phonons are, however, not necessarily topological. The coupling between chiral and topological phonons with electronic and magnetic excitations can give rise to unusual quantum phenomenon, such as chiral superconductivity, chiral density waves, magnetism, and unconventional transport properties.